\documentclass[twoside]{article}

\usepackage{mathpazo} 
\usepackage[T1]{fontenc} 
\usepackage{microtype} 

\usepackage[hmarginratio=1:1,top=32mm,columnsep=20pt]{geometry} 
\usepackage{hyperref}  

\usepackage[hang, small,labelfont=bf,up,textfont=it,up]{caption}
\usepackage{booktabs} 
\usepackage{float} 

\usepackage{lettrine} 
\usepackage{paralist} 

\usepackage{abstract} 

\usepackage{titlesec} 
\renewcommand\thesection{\Roman{section}}
\titleformat{\section}[block]{\large\scshape\centering}{\thesection.}{1em}{} 

\usepackage{fancyhdr} 
\title{\vspace{-15mm}\fontsize{24pt}{10pt}\selectfont\textbf{Reichenbach's Transcendental Probability}} 

\author{
\large
\textsc{Fedde Benedictus and Dennis Dieks}\\
\normalsize History and Foundations of Science\\Utrecht University\\
\normalsize \href{mailto:f.j.benedictus@uu.nl}{F.J.benedictus@uu.nl}
}
\date{\today}

\begin{document}

\maketitle

\thispagestyle{fancy}


\begin{abstract}

\noindent
The aim of this article is twofold. First, we shall review and analyse the Neo-Kantian justification for the application of probabilistic concepts in physics that was defended by Hans Reichenbach early in his career, notably in his dissertation of 1916. At first sight this Kantian approach seems to contrast sharply with Reichenbach's later logical positivist, frequentist viewpoint. But, and this is our second goal, we shall attempt to show that there is an underlying continuity in Reichenbach's thought: typical features of his early Kantian conceptions can still be recognized in his later work.
\end{abstract}


\tableofcontents
\setcounter{tocdepth}{2}

\section{Introduction}

The standard story of how in the beginning of the 20\textsuperscript{th} century the Positivism of Comte and Mach was transformed into Logical Positivism was challenged by Michael Friedman as early as 1983 \cite{f1983}. Against the account according to which positivist philosophy was simply extended into fields of study other than Comte's sociology and Mach's physics, and augmented (e.g., with the `verifiability theory of meaning'), Friedman pointed out that Logical Positivism in its early stages was thoroughly influenced by Neo-Kantian philosophy. Hans Reichenbach's philosophical development is a case in point. As discussed by Friedman, in 1920 Reichenbach wrote a book in which he made an explicit effort to reconcile Neo-Kantianism with modern science, in particular with the then revolutionary special and general theories of relativity. It was only some time later, after an exchange of letters with Moritz Schlick, one of the leading figures of the Logical Positivist movement, that Reichenbach converted to Logical Positivism. Schlick had criticised Reichenbach for unduly clinging to an outdated philosophical system that made use of the Kantian concept of the \emph{a priori}---albeit in a `relativised' form. Schlick objected that `relativised a priori' statements were not really a priori in the Kantian sense at all but should rather be seen as \emph{conventions}. Reichenbach was swayed by Schlick's argumentation and from the time of this exchange refrained from using terminology involving the Kantian a priori and instead started using the term convention; and in his later work he severely criticised the rigidity of the Kantian system.

However, in this article we shall attempt to show that underneath this undisputed change in Reichenbach's attitude there is a remarkable continuity in his work that goes back to even his earliest ideas. To start with, Reichenbach's work of 1920 makes essential use of ideas that come from his 1916 dissertation---something which has not yet received the attention it deserves. Consider, for example, the following words of Friedman:

\begin{quote}
``[i]t is in no way accidental that coordination as a philosophical problem was
first articulated by scientific philosophers deliberately attempting to come to
terms with Einstein's general theory of relativity. Indeed, Reichenbach in 1920,
together with Moritz Schlick in virtually contemporaneous work, were the first
thinkers explicitly to pose and to attempt to solve this philosophical problem.''
(\cite{f2001}, p78)
\end{quote}

However, already in 1916 Reichenbach emphasized that ``physical knowledge consists in the coordination of mathematical equations with particular objects of empirical intuition''(\cite{r1916}, p123) and investigated the details of this coordination, in particular the requirement of `uniqueness'. As far as evidence goes, Reichenbach was not yet familiar with Einstein's relativity in this period. As mentioned, this early work of Reichenbach falls squarely within the Kantian tradition (as can be seen from the expression `empirical intuition' in the above quote). Coordination played an important role in Reichenbach's work both in 1916 and in 1920, and Reichenbach's proposals of 1920 can be seen as a natural extension of his 1916 ideas. But as we shall argue, also after Reichenbach's `conversion' to logical positivism there is a thorough-going continuity, even though some of the philosophical labels used by Reichenbach changed.

This continuity appears clearly when we analyse the role of a concept that has played an important part throughout Reichenbach's scientific career, namely the concept of probability. In the next section we shall discuss the Neo-Kantian background against which Reichenbach wrote his PhD-thesis about this topic, after which we shall take a closer look at this thesis itself (section 3). In the fourth section we shall argue that Reichenbach's early concept of probability formed an important part of the conceptual background from which Reichenbach undertook to save Kantian epistemology in the face of Einstein's relativity. Reichenbach's turn away from Kantianism---his works in the 20's and 30's---and the continuity with his earlier Kantian work will be discussed in section V.

\section{Background}

\subsection{Epistemology} \label{epist}

During the 19\textsuperscript{th} century Kantian philosophy had been in and out of vogue; from the 1860's onwards a new wave of interest in Kantian thought arose (\cite{waarstaatdatdan?}). When Reichenbach's dissertation was published in 1916, its philosophical background was this intellectual climate of Neo-Kantianism. To properly understand what motivated the Neo-Kantians it is helpful to remind ourselves of the leading ideas of Kantian philosophy; for our purposes it suffices to look at Kant's views about scientific epistemology.

Kant asked himself the \emph{transcendental question}: `what conditions must be fulfilled in order to make knowledge possible at all?' `What features must the world, as given to us in our scientific theories, have in order to make these theories viable?' According to Kant there are several characteristics that reality must have in order for knowledge to be possible. For the sake of illustration, consider an arbitrary physical object. We cannot help conceptualise this object as something extended in space and placed in time. That is, even before we start any empirical investigation---even before we have ever seen an object---we can convince ourselves that we can only know objects as being placed in space and time. Kant accordingly held that certain elements of our knowledge are \emph{a priori}: we can know that they must be part of our knowledge even prior to observation, on the grounds that they are necessary to make knowledge possible. These a priori elements in a sense \emph{constitute} the objects, or more generally the things our knowledge is about. Not in the sense that we do not need to do empirical investigations to know their specific characteristics, but rather in the sense that their general nature (like being spatiotemporal) can be established beforehand.

In the century following Kant, Neo-Kantians elaborated on the exact role of these a priori elements in our knowledge. On the one hand they held on to the idea that certain structural features must necessarily be in place a priori to make knowledge possible. On the other hand, they tended to be more liberal than Kant himself concerning the apodictic character of the a priori. Kant had held that since a priori elements are a \emph{sine qua non} for our knowledge, they cannot be contradicted by experience. Experience is itself partly determined by the a priori elements, so stating that experience can serve to adjust these elements appears contradictory. Kant concluded that the a priori elements cannot be changed: they are immutable characteristics of human knowledge. This immutability was called into doubt among the Neo-Kantians, however. One famous motivation for this doubt was provided by the status of Euclidean geometry as an a priori element of our knowledge: Kant had associated Euclidean geometry with the structure of our intuition and our possibilities of visualising, so that it became a precondition for the possibility of spatial knowledge. When it was established in the 19\textsuperscript{th} century that non-Euclidean geometries were serious mathematical contenders, this was a reason for some philosophers to consider the possibility of changes in the a priori framework---a line to be followed by Reichenbach as we shall see.

\subsection{Probability}

The theory of probability is a late addition to mathematics. Whereas geometry received its first axiomatisation already in the fourth century B.C., it took almost another two millennia for a more or less systematic and coherent interpretation of the theory of probability to crystallise. This interpretation, which has become known as the classical interpretation, had been implicit in the work of earlier scholars (e.g. that of Galileo [1620]), but its foundation is traditionally placed in the second half of the 17\textsuperscript{th} century, in the exchange of letters between Blaise Pascal and Pierre de Fermat. Their seminal work was continued and systematized by Pierre-Simon de Laplace in his ``Th\'eorie analytique des probabilit\'es'' [1820].

Laplace was not always completely clear and consistent about whether he believed statements of probability referred to reality itself or merely to our (deficient) knowledge of reality, but a 'Principle of Insufficient Reason' is uncontroversially essential to his use of the concept of probability. According to Laplace's classical definition, the probability of an event is equal to the proportion of the number of favourable cases to the total number of possible cases; the principle of insufficient reason is used to determine the equivalent possible cases. We assume equipossibility\footnote{The concept of equipossibility is closely related to, but not identical with, equiprobability. The recognition of this distinction is important, for it allows Laplace to define probability in a non-circular way.} if we have no reason to prefer any one of them over the others. In his more systematic passages Laplace makes it clear that according to him there is nothing in probability that contradicts determinism. He famously speculated about an entity with infinite powers of calculation in a perfectly deterministic world: for such a hypothetical creature---`Laplace's demon'---the laws of probability are of no value. This thought-experiment illustrates that for Laplace, at bottom, it is only \emph{our ignorance} about details that legitimates statements of probability.

\section{Reichenbach's 1916 Dissertation: ``The Concept of Probability''}

Being inclined towards analytical thought, Hans Reichenbach (1891-1953) started a study of civil engineering at the Technische Hochschule in Stuttgart in 1910. He quickly tired of this, however, because his interests were mainly theoretical. His interest in philosophy was not too great at this time: he felt a certain disdain for its inexactness. Furthermore, with the exception of the writings of Kant there seemed to be no clear connection between philosophy and the natural sciences. In an autobiographical sketch he wrote 30 years later, Reichenbach tells us that despite his disdain for philosophy, he had always been interested in the philosophical foundations of the kinetic theory of gases. This interest led him to a concept he deemed fundamental to these foundations, the concept of probability.


Particularly the issue of whether, or to what extent, the laws of probability can be said to yield a genuine description of reality fascinated Reichenbach, which motivated him to choose this question as the topic of his PhD-dissertation: "The Concept of Probability in the Mathematical Representation of Reality" (Der Begriff der Wahrscheinlichkeit f\"ur die mathematische Darstellung der Wirklichkeit) [1916].

Reichenbach starts his PhD-dissertation with an appraisal of the philosophy of his time: he observes that the discussion of foundational issues has split the community of philosophers in two. The discovery that in all knowledge there are elements of `subjectivity' (in the Kantian sense) has led many to believe that objective knowledge is an utter impossibility. Others have seen these elements as a reason to change the very aim of philosophy: those philosophers wish to delineate exactly what the structure of knowledge is, which components are coming from us and which can be said to originate outside of us.
	It is in this vein that Reichenbach introduces his analysis of probability. As he says, an unanalysed notion of probability may be sufficient for dealing with the irregularities of daily life, but the great importance of this concept for the exact sciences demands a rigorous epistemological analysis.

Reichenbach sets the tone of his investigation by noting that many philosophers, impressed by determinism, share the misconception that probability is only an expression of our personal expectation:

\begin{quotation}
"this .... has led many philosophers to believe that the concept of probability only represents our subjective expectation, which does not have any connection to the real world" (\cite{e1916}, p41).
\end{quotation}

To develop his own, sophisticated view about the relation between physical reality and probability, Reichenbach compares and contrasts his ideas with those of two important contemporary thinkers about probability, Carl Stumpf and Johannes von Kries.

\subsection{Carl Stumpf}

In order to illustrate where the subjectivist account goes wrong, Reichenbach turns to the work of Carl Stumpf. Stumpf had adopted Laplace's definition of probability with only small adjustments:

\begin{quotation}
"We say that a certain state has a probability of $n/N$ if we can regard it as one of $n$ favourable cases within a total of $N$ possible cases, of which we know only one is true, but we don't know which."
(\cite{Stumpf1892a}, p48)
\end{quotation}

A simple example is a throw with two dice. If we ask ourselves what is the probability of 'snake-eyes' (two ones), we have to realize that there are 36 possible outcome situations, of which only one realizes one-one and is therefore `favourable'. Therefore, the probability is one in 36. Reichenbach, however, considers a different example to show the weakness of this definition. Suppose we know of a comet that is in a stable orbit around the sun, and we ask ourselves what the probability is that this orbit has the shape of an ellipse. Now there are four possible cases - the shape is either a hyperbola, a parabola, a circle, or an ellipse - of which only one is favourable, so the chance has to be one in four. But what if we suddenly realise that the circle is nothing but a special case of the ellipse (the limiting case of the eccentricity going to zero)? This would leave only three possible cases, so the probability would now become one in three. Reichenbach takes this as a sufficient ground to reject the subjectivist approach: such dilemmas (comparable to Bertrand's paradox) do not have an objective solution in the subjectivist approach. As Reichenbach puts it, Stumpf's definition cannot deliver what we should expect from a scientific notion of probability: it does not provide us with a basis for objective and rational expectation.

\subsection{Johannes von Kries}

For Reichenbach, Johannes von Kries exemplifies the opposite stance regarding probability. In Reichenbach's eyes, von Kries' approach represents an advance, but it does not lead us far enough: on scrutiny, it turns out not to achieve real objectivity.

Von Kries condemns the use of the Principle of Insufficient Reason. To replace it with something more objective, he invokes the notion of \emph{event spaces} and a rule to assign a measure to them. Consider the example of the roll of a die again. There are six different outcomes, i.e.\ situations in which different sides of the die face upwards after it has been thrown, and these situations constitute a universe of events. But these events can be split up further: each outcome can be realized by very many microscopic configurations that all manifest themselves macroscopically as the same outcome. Now the probability of an outcome can be calculated if we are able to somehow assign a measure to this `space' of more fundamental microscopic events. This thought leads von Kries to define the probability of a specific outcome as the ratio of the size of its specific event space to the size of the space of the totality of all events, corresponding to all possible outcomes. It is then possible to say which events are equipossible---namely the events whose event spaces are of equal size. Given the symmetries of the situation, this procedure would plausibly lead to assigning equal probabilities to all possible results of the throw of a fair die: one in six.

\subsubsection{Elementary Event Spaces}
But what if our die is biased, loaded in such a way that the sides are not equipossible? In this case the event spaces corresponding to different outcomes cannot be of equal size. Von Kries argues that in such cases we should trace back the causal nexus until we find event spaces that are equal in size and do not change their sizes anymore if the causal nexus is traced still further back. We may illustrate the idea by considering the biased die in more detail.

Since the die is biased the absence of equipossibility can be observed (the possible outcomes do not occur equally often), so we start tracing back the chain of causes, going to a more fundamental level of description. Given the asymmetries in the physical situation---the die is loaded---we shall find that the number of initial states, in a micro description, that leads to one outcome will be different from the number of initial conditions that results in another outcome (we assume that the mechanism of casting the die is the same in all cases). When we go back even further in the causal nexus, at some point these initial micro states will not split up any further, at least not in an asymmetric way. If this is the case, von Kries dubs the events `elementary'. The basic principle of von Kries's method is to take these elementary events as equipossible. So we have to go on with our causal analysis back in time until we find `simple, non-composite' causes; these event spaces we ultimately arrive at should possibly be defined in terms of the states of the individual atoms of which the die consists.\footnote{the atomic concept had not yet gained general acceptance in von Kries's time, but `small quantities of matter' could serve a similar explanatory role.}

So for every chance event there are elementary event spaces, but not every possible outcome necessarily has the same number of elementary events associated with it. If, in the case of the biased die, we ask for the chance of six eyes facing upwards after it has been thrown, we should consider the totality of possible microscopic configurations ($N$), and determine the number of these that will manifest themselves as the desired outcome ($n$). The sought-for probability is then the ratio $n/N$.

We are now in a position to define an objective notion of equipossibility: those events are equipossible whose elementary event spaces consist of an equal number of elementary events. As the elementary event spaces are the same for every observer, independent of his state of knowledge, von Kries concludes that his notion of probability is fully objective. Von Kries demands that equipossibility has its roots in physical facts---it is due to a regularity present in nature, not in a subjective degree of knowledge. Not only are the elementary event spaces the same for all observers, they also directly reflect the structure of reality.

\subsection{Reichenbach}
Reichenbach criticises von Kries' approach on two counts. First, Reichenbach says, the `principle of event spaces' is problematic. According to Reichenbach the characterisation of elementary events and their probability ultimately boils down to a principle of insufficient reason that is similar to that of Stumpf or Laplace. Although von Kries has given a characterisation of his elementary event spaces in physical, causal terms, Reichenbach objects that von Kries has not given us a compelling justification for translating the equal size of these elementary event spaces into equality of chances. How do equal elementary event spaces lead to equipossibility? In answering this question von Kries seems to make use of the Principle of Insufficient Reason after all: if there are elementary event spaces of equal size we have no reason to favour one of them over the others, so we just assume that the events with which they are associated are equally possible. The determination of equipossibility thus depends on our state of knowledge, albeit in a hidden way, so that this supposedly objectivist view is actually not objective at all. An element of subjectivity, relating to our knowledge, has crept in. Therefore, Reichenbach concludes, von Kries' approach should have no place in objective science.

A closely related objection that Reichenbach formulates is that von Kries does not deduce his principle of event spaces from a systematic conceptual framework---von Kries merely postulates it. Reichenbach concludes that von Kries's principle must be replaced by one that can be justified from an encompassing scientific philosophy.

\subsubsection{The Probability Function} \label{pf}
As we have seen, the objectivity of any account of probability hinges on how this account establishes equipossibility. Reichenbach agrees with von Kries that statements of probability should be grounded in objective symmetries. But how can equipossibility be inferred from physical symmetries without using the Principle of Insufficient Reason?

To start answering this question Reichenbach makes a typically Kantian move: he argues that we first of all \emph{must assume} that \emph{a probability function exists} in order to be able to make sense of experience at all. Without the assumption that the irregular data coming from experiments represent underlying constant (probability) values, no knowledge of reality would be possible.

In the next section we go into the question of how the existence of such a probability function can justify the equipossibility of certain events, but let us first look at the assumption itself and the ideas behind it. Consider a roulette-like game: we have a rotating disc with alternatingly black and red segments and a fixed pointer next to the disc.\footnote{Reichenbach uses a somewhat different `probability-machine' in his dissertation.} The disc is given a swing with a fixed amount of force, and when it has stopped rotating the pointer indicates either a red or a black segment. In an ideal deterministic setting the colour indicated by the pointer will always be the same in repetitions of the experiment, if the amount of force that is used is fixed. But in real experiments this will not happen. There are always myriads of perturbing factors (e.g.\ the wind, seismic activity, instability of the hand of the croupier, the gravitational pull of a passing comet, and so on). Therefore, the outcomes will not always be the same, but will vary. We can represent the results of the game in a bar chart, in which the relative frequencies with which the segments of the disc are indicated are represented.

Now, it is not self evident, analytic or a matter of logic that in repetitions of the experiment this bar chart will eventually converge to a stable function. Logically speaking, we could be living in a chaotic universe in which no conclusions about probabilities could be drawn at all from an experimentally produced bar chart: the relative frequencies could go on fluctuating without approaching any limit at all. But in such a world no knowledge coming from statistical data would be possible. In other words, if scientific knowledge is to be possible at all, we have to adopt the \emph{a priori} principle that the bar chart is representative of a probability distribution. In fact, we have to assume this probability distribution to be a smooth function in order to come to quantitative results, as will be explained in the next subsection. This \emph{a priori} assumption of the existence of a continuous probability distribution is the essential ingredient of Reichenbach's 1916 transcendental probability theory.

\subsubsection{Equipossibility Revisited}

In the roulette-like game we just described we would intuitively expect that the ratio of the black and red areas determines the ratio of the probabilities of black and red, respectively. In the physically symmetrical situation of equal red and black areas we expect equipossibility of red and black. As we have seen, von Kries attempted to justify this intuitive probability judgment by considering elementary event spaces and assigning them equal probabilities. Let us go along with this idea to some extent, in spite of the weaknesses of von Kries's justification, and let us think of the disc as being divided up in very many infinitesimally narrow segments, each one either black or red. Now, if we assume that a continuous probability distribution exists, defined over these infinitesimally narrow segments, it follows mathematically that adjacent segments (infinitesimally close to each other) have the same probability. The assumption of the existence of a continuous probability distribution thus leads to an immediate improvement on von Kries: we can now \emph{prove} that certain elementary events are equipossible.

Consider the symmetrical situation in which the red and black segments alternate: after each infinitesimal red segment there is a black one, and so on. In this case the adjacent black and red elementary events are equipossible\footnote{Actually, the probability function need only be integrable to reach this conclusion. Reichenbach acknowledges this, yet sticks with the assumption of continuity.}. Physical symmetry is thus translated into equipossibility. In more complicated situations, in which the black and red segments do not cover equal areas on the disc we can generalize this argument\footnote{See (\cite{butterfield}) for a recent discussion.}.
The assumption of the existence of a continuous probability function therefore gives access to equipossibility judgments that do not depend on our knowledge. Objective symmetry can lead, via the existence of a continuous probability function, to the assignment of equal probabilities.

The continuous probability distribution itself cannot be observed---no matter how often we repeat the experiment, we always end up with a step-wise bar chart as the result of measuring relative frequencies. The existence of the probability distribution is therefore not given in experience, but is a \emph{precondition} of (a certain kind of) experience: without it we could not have experimental knowledge of probabilities.

Once Reichenbach's probabilistic \emph{a priori} is accepted, this allows us to see observed relative frequencies as representative of probabilities and therefore to extrapolate probabilities from observed frequencies. This provides the link between reality and probability, and it is the background of Reichenbach's claim to have removed the last non-objective element from the concept of probability. We no longer have to use a principle that rests on lack of knowledge, we only have to assume the existence of a continuous probability function.

Two remarks to end this subsection. First, it should be clear that nothing has been claimed about the precise form of the probability function, except its continuity. Second, it is important to realise that Reichenbach's approach differs in an important way from frequentism. Frequentists \emph{define} the probability of an event as the limit of its relative frequency in repeated trials---the probability therefore exists only \emph{a posteriori}, if the limit relative frequency is established to exist inductively. However, Reichenbach assumes the existence of a continuous probability function \emph{a priori}, as a \emph{precondition} for the possibility of knowledge. This assumption makes it possible to exploit statistical data for making probability statements, without entering into questions about the convergence of relative frequencies in an infinite series of experiments.

\subsubsection{The Transcendental Deduction} \label{trded}

Reichenbach justifies the assumption about the existence of a continuous probability function in typically Kantian fashion, by providing a `transcendental deduction' of it. In his dissertation Reichenbach describes the Kantian system as follows (\cite{r1916}, p110).

Kant importantly distinguished mathematical from physical judgments. Consider, for example, an experiment in which a stone is dropped from a tower: if the fall of the stone is described purely mathematically, the objects to which this description refers are mathematical in nature---a perfect sphere moving along a perfectly straight line in an absolute vacuum. The truth of such a description can be judged by pure intuition alone---the relations between the objects are inherent to the concepts that are used. This differs in an essential way from physical statements, whose truth can only be judged with the help of observation. The object of physical knowledge is not a mathematical perfect sphere, but the nicked and notched real stone, or the imperfect orb of the moon.

A central role in Kant's philosophy is played by deductive arguments that show the necessity of specific a priori concepts. In Reichenbach's version, these `transcendental deductions' serve to ground the assignment of particular mathematical concepts to empirical reality. The possibility of physical knowledge depends on this assignment, without which the empirical (physical) world would be without form. However, Reichenbach emphasizes,  since there are no closed systems in reality, physical objects as they are given to us in experience are always subject to disturbances and deviate in countless ways from their mathematical conceptualisations. Therefore, the possibility of physical knowledge depends crucially on the possibility of making systematic approximations to ideal mathematical structures.

In his ``Critique of Pure Reason'' (``Kritik der Reinen Vernunft'', [1783, \cite{kritik}] Kant had already discussed the concept of causality and assigned it the status of a \emph{synthetic a priori} concept. Although the principle of causality is beyond empirical justification, the very possibility of physical knowledge (which is about causes and their effects) requires either a principle of causality or a principle of lawful connection. However, Reichenbach points out that only in an ideal physical situation (one in which observations are not influenced by any external disturbances) a principle of lawful connection by itself suffices to fix the objects of our knowledge. In the roulette-like game, such a situation would correspond to a very monotonous game: if the disc is swung with a fixed amount of force, the outcome of the game would always be the same, without any variation.
In all \emph{real} physical situations an additional principle, something more than lawful connection, is needed to transform observations into knowledge. In the presence of disturbances, following just the principle of lawful connection would lead us to conclude that the experiment was about a different object in each individual case. Therefore, the Principle of Lawful Connection needs to be supplemented in order to accommodate that the same physical situation may give rise to a distribution over outcomes---so that these different outcomes can provide information about the same object of knowledge. This leads Reichenbach to conclude that besides a principle of lawful connection, a principle of lawful distribution must be assumed in order to make sense of the possibility of robust knowledge in the face of irregularity and randomness.

The just-described justification of the principle of lawful distribution has the typical form of a Kantian transcendental deduction: it analyses the conditions that must be fulfilled in order to make knowledge possible. The conditions that are arrived at (in this case the need to assume the existence of a probability distribution that is characteristic of a given experimental set up) have the status of synthetic a priori principles.

Obviously, the applicability of this argument is not restricted to games of chance. In fact, \emph{all} measurement outcomes in physics require the existence of a continuous probability function if they are to serve as a basis of knowledge. Indeed, suppose we want to talk scientifically about some object; then we shall want to make quantitative statements about this object, based on measurements. But if we repeat the measurements, the second measurement will never yield exactly the same result as the first. To be able to formulate a scientific hypothesis about our object, we need first of all to justify our belief that our repeated measurements indeed pertain to one and the same object; and for this we need the principle of lawful distribution.

The inaccuracy of everyday measurements ensures that this complication has no consequences for our daily lives. Science, however, does not get off that easily. Confirmation of hypotheses would never be possible if we refrained from accepting the existence of a probability distribution. To do science we must assume that our measurement outcomes may deviate from what is dictated by strict causal rules, and we must assume that the outcomes follow a specific distribution function.

\section{1920: Neo-Kantian Epistemology and the Theories of Relativity}


Four years after the publication of his PhD-dissertation Reichenbach returned to Stuttgart, where he had studied civil engineering some ten years earlier. In Stuttgart he became an instructor in physics and eventually associate professor. It was here that Reichenbach wrote his habilitation thesis, on the relation between the epistemology of Kant and the newly developed theories of relativity: ``The theory of relativity and a priori knowledge'' \cite{r1920}.


The book starts with identifying those elements in the theories of relativity that contradict Kant's theory of knowledge. One of the a priori elements of pure intuition that Kant had identified was that of absolute time, with absolute simultaneity as an essential ingredient. When Reichenbach was confronted with the theory of special relativity \cite{e1905} (in all likelihood only after he had written his PhD-dissertation), he faced the challenge of reconciling the new relativistic concept of time with Kantian doctrine. Indeed, in Einstein's theory there is no absolute simultaneity: two observers will disagree on the simultaneity of distant events if they are in motion with respect to each other.


In the Introduction we already noted a similar problem relating to the development of non-Euclidean geometries. This problem was also aggravated by the theory of relativity: according to the special theory of relativity Euclidean geometry will generally cease to be valid in accelerated frames of reference, and Einstein used this fact to argue in his General Theory \cite{e1916} that gravity will manifest itself, among other things, through the presence of non-Euclidean geometrical relations.

\subsection{Uniqueness of Coordination}\label{uniqueness}


As we pointed out, in 1916 Reichenbach had already restated Kant's transcendental question in terms of the relation between mathematical concepts and physical knowledge and had tried to clarify the precise nature of the a priori in these terms. The knowledge acquired in physics, according to Reichenbach, consists in a coordination of the objects of experience to mathematical objects---a physical judgment asserts the validity of using a certain mathematical structure for describing reality. An example of such a coordination is the coordination of geometrical concepts to the space-time continuum in which we live. This makes it possible for Reichenbach to apply his earlier ideas about coordination directly to the new situation created by the theories of relativity.


The coordination between mathematics and reality starts from certain principles, which Reichenbach calls \emph{coordinating principles}. Within an epistemological system that depends on such a coordination, the possibility of unambiguous scientific knowledge hinges on the attainability of \emph{uniqueness} (Eindeutigkeit) of the coordination. In Reichenbach's own words: ``truth [is defined] in terms of unique coordination'' (\cite{r1920}, p41). The knowledge-coordination is unique if every element of reality is uniquely assigned its own element of the mathematical system that is used, and is thus uniquely defined. Everything physical (according to our conceptual framework) can then be objectively described in terms of mathematical concepts. However, to simply assume the uniqueness of any system of coordination would be assuming that there exists a pre-established harmony between reason and reality. To make such an assumption would be stepping outside the boundaries of empirical science. We therefore have to verify whether the coordination that we use indeed leads to a unique description of reality. That means that if we use a certain conceptual structure to describe physical reality, we have to make sure that the various elements occurring in that structure receive an unambiguous characterisation, even if we acquire information about them via very different methods. Uniqueness of coordination manifests itself in a convergence of different experimental techniques. For example, Avogadro's number can only be considered an objective and unique element of physical reality because radically different methods of determining it (e.g., via the theory of gases in equilibrium and via fluctuation phenomena) lead to \emph{essentially} the same result. The qualification `essentially' is crucial here: as we have seen, we should not expect perfect agreement of observations even if these pertain to the same object of knowledge. This imperfect agreement can nevertheless be reconciled with a unique coordination if we are allowed to extrapolate and `correct' experimental data. We must be allowed to interpret small differences in our measurement outcomes as mere fluctuations, deviations from a constant `real' value. That is, we must be allowed to regard the variations in question as representative of a probability distribution, so that we can extrapolate from average values to ideal mean values that stand for the undistorted physical quantities. This procedure is called \emph{normal induction}, says Reichenbach. He emphasizes the importance of this principle:


\begin{quote}
``The principle of normal induction, above all other coordinating principles, is distinguished by the fact that it defines the uniqueness of the coordination.'' (\cite{r1920}, p64)
\end{quote}


The theories of relativity have serious consequences in this regard, for they seem to imply that different methods of experimentation do not always lead to the same results. First it was the theory of special relativity which presented a problem for the achievement of uniqueness. According to this theory observers in rapid motion relative to each other disagree widely on the simultaneity of spatially separated events. Therefore, this theory tells us that we end up in contradictions if we hold on to the way in which we usually extrapolate our experimental data---within the conceptual framework of Newtonian physics. Matters become even worse for the Kantian doctrine if we take the consequences of the theory of general relativity into account. In this theory not only simultaneity depends on the observer's state of motion: geometrical relations are affected in a similar way. According to Einstein's principle of equivalence, the presence of a gravitational field can be seen as a consequence of being in an accelerated frame of reference. This means that the conclusion from the special theory that a non-Euclidean geometry applies to accelerating frames carries over to general relativity: in the presence of gravitation, geometrical relations will generally be non-Euclidean. However, in a freely falling frame geometrical relations will be Euclidean again. In Einstein's theory Euclidean geometry therefore loses its preferred and absolute status, as the geometry can change by the presence of gravitating bodies and by the transition from one frame to another.


Such considerations show that Kant's original epistemology is contradicted by the theories of relativity in more than one way. Combining this with the fact that the theory of general relativity has received ample experimental verification, we are tempted to conclude that the Kantian a priori elements are in need of adjustment. Should we be pragmatic and freely choose completely different principles---ones that would permit a unique coordination in a simple way? Such a drastic conceptual reform would conflict directly with Kantian thought and Reichenbach is not prepared to go along with it. As he comments:

\begin{quote}
``it is a moot question and irrelevant for Kant whether some day reason will change because of internal causes. [...] All that his theory excludes is a change of reason and its order principles by \emph{experience}: `necessarily true' must be understood in this sense.''(\cite{r1920}, p54)
\end{quote}

In other words, although it might be conceivable that a Kantian will one day adopt modified a priori principles as a result of changes in human reason and intuition, to do so on the basis of the outcomes of experiments is in conflict with the very spirit of the Kantian system.

\subsection{Gradual Extension}


To resolve the tension between Kantian epistemology and the theories of relativity, Reichenbach now makes a distinction between two ways in which the `a priori' functions in Kant's theory of knowledge. Firstly, the original Kantian a priori is characterised by \emph{apodicity}: the a priori elements are necessary for knowledge in an absolute way, they are immutable because as preconditions of experience they cannot be contradicted by experience. The second characteristic of the a priori is that a priori elements \emph{constitute} the concept of object: only through the coordination of concepts to physical reality do physical objects become well-defined at all.


This distinction points the way to how the dilemma of the previous section can be removed, namely by abandoning the apodictic a priori and thus by \emph{partly} weakening Kant's original conception. In order to counter the objection that this will lead to contradictions, as the a priori is needed to define experience itself, Reichenbach now invokes his earlier ideas about probability and approximation; this will make it possible to retain the constitutive part of the a priori.

If the a priori is no longer apodictic it can in principle change. But Reichenbach does not envision arbitrary, drastic and freely chosen changes. He introduces the concept of  \emph{gradual extension} (\emph{stetige Erweiterung}): instead of assuming that the a priori elements will always remain the same, we should allow experience to guide us to \emph{ever more accurate} principles of coordination, bringing us ever closer to a truly unique coordination. Only this principle of gradual extension makes the notion of a `relativised a priori' consistent\footnote{\cite{f1999}, p44.}: although it is still true that we need a priori concepts in order to make sense of physical reality, we do not need to assume that these concepts are completely rigid. Our coordinating principles, and the nature of our concepts, may evolve with the development of our knowledge. In spite of such conceptual changes, the older constitutive system keeps its value in structuring our experience, since the new concepts are \emph{refinements} of the old ones, found in a process of continuous generalization.


This change in outlook depends in a fundamental way on probability theory and on the principle of normal induction that we encountered earlier. If we assume (as in section \ref{trded}) that the outcomes of our measurements vary and that the true values must be inferred by a normal inductive procedure, the flexibility of this procedure enables our measurements to support both the old coordinating principles as well as slightly modified new coordinating principles. For example, the discrepancies between Newtonian absolute simultaneity and special relativistic simultaneity are in everyday circumstances so small that they are insignificant and disappear against the background of the spread in actual measurement results. If a principle of normal induction were not in place, we could not consistently use different conceptual systems next to each other, and could not view earlier theories as approximations to later ones (in the sense in which Newtonian mechanics can be viewed as an approximation to relativistic mechanics).


Thus, the theories of relativity on the one hand prove the original Kantian epistemology to be inadequate but on the other hand also show the way towards a `relativised' epistemology. If we replace the demand of a rigid uniqueness---without approximations---in the coordination of mathematics to reality with the demand that a principle of normal induction is valid then it becomes possible for new theories to replace earlier theories, even if their concepts are different.


Superficially it might seem as if ``The theory of relativity and a priori knowledge'' represents a firm step away from Reichenbach's earlier Kantian epistemology: the Kantian concept of the a priori is explicitly decapitated and left exposed to experimental verification. But such an assessment would overlook the essential similarity between Reichenbach's 1916 work and that discussed in this section. Not only does the idea of knowledge as a coordination of mathematics to reality originate in Reichenbach's dissertation, the most important concept in this dissertation is also of vital importance for the coherence of the epistemological system put forth in 1920: the principle of normal induction, which allows us to interpret (small) variations in the outcomes of our measurements as insignificant---this principle makes it possible to refer to a unique object in spite of non-uniqueness of measurement outcomes, and also allows us to combine different conceptual frameworks with the same empirical data. The a priori principle of the existence of a continuous probability function is of primary importance here, as Reichenbach himself points out. When he says, in 1920, ``if in spite of the inexactness of any measuring device a unique coordination is assumed, the principle of normal induction must be retained'' (\cite{r1920}; 1920, p64), he follows this up with a footnote in which he refers to his 1916 dissertation for a justification of this statement.


Reichenbach's principle of the existence of a continuous probability function, which grounds his doctrine about the way probability applies to reality, thus fits in naturally with the rest of his Neo-Kantian methodology. The procedure of normal inductive inference that hinges on this principle plays a crucial double role in Reichenbach's epistemology: not only does it serve to demarcate the individual physical object, it also enables a gradual extension of our principles of coordination.

\section{Reichenbach's Later Views on Probability and Induction}


\begin{quote}
``Even though [the] first among my papers referring to the problem of probability was written under the influence of Kant's epistemology, it seems to me that the result concerning the theory of probability can be stated independently of Kant's doctrine and incorporated in my present views.'' (\cite{r1949}, p355)
\end{quote}


In 1949 Reichenbach published his mature views on probability and induction in \emph{The Theory of Probability}---a book written long after his renunciation of Kantianism\footnote{\emph{The Theory of Probability} was Reichenbach's own translation (with which he was assisted by E.H. Hutten and his wife Maria Reichenbach) into English of his 1935 work \emph{Wahrscheinlichkeitslehre}. The revised version differed significantly from the original publication, which motivated the publisher to style it a new `edition' rather than a new `print'.}. Nevertheless, on closer scrutiny it turns out that several motifs that were characteristic of his early Kantian approach can be identified in this later work as well. It turns out that there exists a strong underlying continuity in Reichenbach's thought, which may appear surprising given his `conversion' to logical positivism.

In 1949 Reichenbach has become a frequentist: he now defines every degree of probability as the \emph{limit of a frequency} within an infinite sequence of events(\cite{r1949}, p68). Of course, two well-known problems immediately present themselves if this definition of probability is adopted. Firstly, not every event to which one wants to assign a degree of probability belongs to a sequence of events. How can one speak of a frequency within a sequence if the event whose probability is considered is a singular event? The second problem arises as a consequence of Reichenbach's demand, which he had emphasized ever since his official abandonment of Kantianism, that empirically meaningful scientific statements must be verifiable. How can a statement ever be verified if it says something about an \emph{infinite} sequence? As only finite sequences can be an object of observation, probability defined in terms of limiting frequencies seems to lead to statements that are not empirical.

\subsection{Posits}\label{posits}


To solve the first of these problems Reichenbach introduces ``a logical substitute that can take over the function of a probability of the single case without being such a thing in the literal sense''(\cite{r1949}, p372). If we want to make a statement about the probability of an event when it is impossible to indicate a sequence to which this event belongs, we should---Reichenbach argues---understand the statement not as an assertion, but as a \emph{posit}. Reichenbach compares this situation with the placement of a bet in a horse-race: we do not say by placing a bet that it is true that the horse will win with a certain frequency, but we act \emph{as if} it were true by staking money on it. Likewise, a posit is a statement which we treat as being true, although its truth-value is unknown. Making posits enables us to define the probability of singular events, by making use of the concepts of an infinite sequence and a limiting frequency in it, even though this infinite sequence does not actually exist. The frequencies and their limits thus play a role on the conceptual side, but do not immediately reflect properties of what we empirically ascertain.


The problem of verifiability is closely related to what we just discussed. In Reichenbach's frequency interpretation a probability statement is a statement about an infinite sequence, whereas any sequence that we can observe is necessarily finite. Direct verifiability of probability statements is therefore obviously out of the question. In order to maintain contact with empirical reality, there is only one way out: we have to justify at least some probability statements by extrapolating from our experimental data. Because the use of an inductive inference is unavoidable if we want to base a probability judgement on empirical data, the problem of the verifiability of probability statements reduces to the problem of the justification of induction (\cite{r1949}, p351). Conversely, the problem of induction is essentially solved as soon as we know how to justify probability statements on the basis of finite frequencies.

The analysis of induction that Reichenbach proposes is perhaps the contribution for which \emph{The Theory of Probability} and its 1935 predecessor \emph{Wahrscheinlichkeitslehre} are remembered best---it has become known as the \emph{pragmatic theory of induction} (see, for example, \cite{skyrms} and \cite{galavotti} for discussion). Importantly, Reichenbach rejects the traditional empiricist project of justifying the validity of induction from what we find in experience; as Hume had already convincingly demonstrated, this project can only lead us into an infinite regress. Instead, what Reichenbach offers is a \emph{vindication} of induction, in the sense of an argument that purports to show that if we make use of induction we have the prospect of making scientific progress whenever this is possible at all: \emph{if} there are in fact universal regularities and limiting frequencies, we shall get close to them if we extrapolate from our finite data. This does not presuppose that the regularities and probabilities are actually there; they may not exist. But in that case the scientific enterprise is hopeless anyway: no scientific knowledge is possible in a world without regularities. In this situation induction will not help us---but then nothing will. On the other hand, the laws and relative frequencies may exist, and it will not hurt to try to find them. Using induction will pay off in the situation in which science \emph{is} possible. As Reichenbach writes (\cite{r1949}, p482):

\begin{quote}
``As blind men we face the future: but we feel a path. And we know: if we can find a way through the future it is by feeling our way along this path.''
\end{quote}

The conditional character of Reichenbach's approach to induction, with its emphasis on the necessity to go forward and accept the associated risks, fits in with his analysis of probability statements as \emph{posits}. Although we cannot observe probabilities and cannot be certain about their existence, we posit them to exist. If we were not to make such a posits, we could not even start applying statistical theory. If we use the symbol $A$ to represent the existence of a limiting frequency, $B$ to represent the use of induction (in the sense of straight extrapolation)  and $C$ to represent the discovery or approximation of the limiting frequency, then Reichenbach's argument can be summarized as follows: $A \to (B \to C)$ (\emph{if} a limit exist, \emph{then} the use of induction will lead us to it).

\subsection{From Kantian \emph{Preconditions of Knowledge} to Posits}


In his 1916 dissertation Reichenbach had argued that we need a Kantian \emph{a priori} in order to be able to apply probability theory to the physical world: we have to assume that the fluctuating results of repeated measurements can be considered as a sample from one continuous probability distribution. Without this assumption we would not be able to attribute the measurement results to an essentially unchanged physical situation, and we would not be justified in speaking, for example, of one and the same object that is present in the repeated experiments. So the assumption of the existence of a continuous probability distribution is a typically Kantian precondition of knowledge: it has to be in place in order to make knowledge possible at all.


The assumption of the existence of a continuous probability function behind empirical data, as discussed by Reichenbach in 1916, is therefore obviously not an inductive result. On the contrary, the assumption that the function exists is needed to give meaning to the measurement results and has an \emph{a priori} status. Once we have the existence of a continuous probability distribution available in our conceptual arsenal, we can make use of it to extrapolate from finite numbers of fluctuating empirical findings to `real values', characterized by the probability function. The same pattern of reasoning applies to induction in general. From Reichenbach's 1916 perspective it is a necessary a priori principle that behind the imperfect regularities that manifest themselves in empirical research there are mathematically well-behaved laws of nature. Actually, as Reichenbach sees it, quite in general one can say that the essence of induction \emph{is} the construction of probabilities from finite data sets.

It is important to note at this point that the \emph{necessity} of the Kantian \emph{a priori} is not completely unconditional, not even in the original Kantian context (as opposed to later neo-Kantian developments). The Kantian synthetic \emph{a priori} judgments must necessarily be fulfilled \emph{if science is to be possible}; in other words, \emph{if} scientific knowledge exists they must be a part of it. This is a necessity with a condition; and the condition could fail to be satisfied. Indeed, logically speaking it could happen that the a priori categories cease to help us out in making sense of the world. This would have the consequence that science would no longer be possible. So, from a formal point of view, a priori judgments have the status of necessary conditions (in the sense this term has in mathematics  and logic), which have to be fulfilled in order to make scientific research sensible at all. This conditional character of the a priori becomes more pronounced in Neo-Kantianism. As we have already seen (sec.\ \ref{uniqueness}), in 1920 Reichenbach explicitly rejects the idea that there is a pre-established harmony between the concepts we intuitively accept and mature scientific theories. That means that the \emph{a priori} concepts and judgments that we use may change in the course of time, as a result of new scientific results. This was precisely what motivated Reichenbach to relativise the a priori in his 1920 book, given the exigencies of relativity theory. At this stage of Reichenbach's philosophical development (his 1920 Neo-Kantianism) the a priori thus acquires the status of a set of concepts and assumptions, suggested and refined by scientific research, which are employed to shape the conceptual framework within which science is conducted. Clearly, this Neo-Kantian a priori is not necessary without restriction, it is hypothetical.


Another preliminary observation to be made concerns the relation between Reichenbach's 1916 (and 1920) philosophy of probability and frequentist interpretations of probability. Although Reichenbach would not call himself a frequentist in 1916, he certainly does subscribe to the idea that probabilities \emph{would} be found as limiting relative frequencies if we \emph{were} able to generate infinite sequences of measurement outcomes. The Kantian twist he adopts makes it possible, however, to regard the actual existence of such infinite sequences, or even very long sequences, as superfluous for the applicability of probability theory. The difference with frequentism (as defended, e.g., by von Mises---more about this below) is thus rather subtle. This difference is first of all that most frequentists would see the applicability of probability theory as an \emph{inductive fact of experience}, based on the consideration of large amounts of statistical data; whereas Reichenbach takes the position that this applicability has to be \emph{presupposed} in order to make the scientific enterprise meaningful.

Let us now turn to Reichenbach's 1949 (and 1935) views on probability and induction, in order to compare them with his earlier ideas. As in his earlier work, Reichenbach in 1949 makes use of the existence of a continuous probability distribution behind actual measurement outcomes. But now this probability is expressly \emph{defined} to be a relative frequency within an \emph{infinite} sequence of events, without the addition of a priori elements to bridge the gap with experience. But of course, we still need \emph{something} to bridge this gap, as discussed in sec.\ \ref{posits}. As we saw, the move Reichenbach now makes is to \emph{posit} the existence of infinite sequences with the right limiting relative frequencies; he proposes to treat statements asserting the existence of these frequencies \emph{as if} they were true, although we refrain from attributing a truth value to them.

The similarity with Reichenbach's earlier Neo-Kantian a priori approach is striking. Both the \emph{a priori assumption} of the existence of probabilities and the \emph{posit} that they exist serve as necessary conditions that have to be fulfilled in order that the application of probability theory makes sense at all. In both cases there is no implication that the probabilities or limiting frequencies actually exist: in the case of the posits this hypothetical aspect is explicitly stipulated, but as we have just argued the same applies to the Kantian framework. Also the Kantian probability function is not assumed to exist unconditionally, but has to be supposed if probabilistic reasoning is to be applicable. But it could always turn out that stable long-run frequencies are not to be had, and in this case we might be forced to abandon the application of probability theory to the physical world---with sad consequences for the possibility of scientific knowledge in general. Clearly, in both his earlier work and in his mature 1935 and 1949 books Reichenbach uses the existence of a continuous probability function as a conceptual tool, intended to justify the application of probability theory to the physical world. We assume this existence, and see how far this assumption takes us. More in general, the same strategy is proposed for all inductive reasoning. To start, we just \emph{assume}, or posit, that there are fixed laws behind the approximate regularities of empirical reality and devise procedures within the conceptual framework defined by this assumption. The assumptions made provide us with a justification for our inductive procedures \emph{within the assumed conceptual framework}. Although this does not give us a justification of induction in an absolute sense, it gives us good reasons to follow an inductive methodology within the `world as conceptualized by us', a well-known Kantian theme.


Reichenbach's later views on probability and induction, and the Neo-Kantianism of his early work are thus much more similar than suggested by the story of Reichenbach's conversion to logical positivism after his confrontation with Schlick. Indeed, an atmosphere reminiscent of neo-Kantianism remains present in Reichenbach's work even where he explicitly criticizes Kant (\cite{dieks2010}). This gives Reichenbach's position a flavour of its own, which distinguishes it from mainstream logical positivism. We already mentioned the distinction between Reichenbach's approach to induction and traditional empiricist attempts to solve the Humean problem. Whereas the traditional empiricist desires to justify induction on the basis of experience alone, and thus faces the problem of infinite regress, Reichenbach breaks this vicious circle by the introduction of \emph{a priori} elements, either in Neo-Kantian fashion or by means of posits. Another significant difference pertains to the status Reichenbach assigned to probability theory. Usually, the interpretation of probability that Reichenbach defended in 1935 and 1949 is referred to as an \emph{frequency interpretation of probability}, suggesting that it belongs to the same category as the interpretations due to Richard von Mises and other empiricists. However, there is at least one important difference between the interpretation of Reichenbach and that of most other scholars within the frequentist school. According to von Mises the theory of probability is a \emph{physical theory}---like mechanics and optics---whereas Reichenbach holds that the theory of probability is a branch of \emph{mathematics}\footnote{Of arithmetics, to be more specific \cite{r1949}, p343.}. This difference is typical of the difference in outlook between Reichenbach and von Mises. Whereas von Mises sees the rules of probability as representative of regularities inductively found in the physical world, Reichenbach thinks of them as a presupposed conceptual system, completely defined and justified within its own framework. In fact, he argues that the probability calculus is nothing but an axiomatic system, which in itself possesses no connection to the physical world. We need ``rules to coordinate an interpretation to the unspecified symbols''; we need ``coordinative definitions''(\cite{r1949}, p40).


This immediately calls to mind the Kantian question as formulated by Reichenbach in 1916 and 1920: the unambiguous coordination of a mathematical system to the physical world. In 1916 Reichenbach already voiced the idea that coordinative definitions are needed to accomplish this. Reichenbach's development of this idea into a coherent epistemological system (with a relativised a priori) in 1920 enabled him to salvage Neo-Kantian philosophy in the face of the theories of relativity: the coordinative definitions define the way we categorize the world (in accordance with Kant's dictum ``percepts without concepts are blind''). Now, in 1949, we see Reichenbach arguing again that coordinative definitions are an essential ingredient needed to link an \emph{a priori} mathematical system, such as the theory of probability, to reality. Considerations about the coordination between a priori conceptual systems on the one hand and physical reality on the other clearly constitute a continuous thread in Reichenbach's epistemology, extending from his early work to the publications later in his life.


The thesis of continuity we have just put forward and explained fits in with a more general reappraisal of the development of Reichenbach's philosophical thought. It may be argued quite generally that Reichenbach's transition from Neo-Kantian epistemology to logical empiricist philosophy should be considered first of all as a change of terminology, with little change in content.  As we know, the now standard account of how Reichenbach changed his views on epistemology from Kantianism to logical empiricism is that after an exchange of letters with Moritz Schlick at the end of 1920 Reichenbach was swayed to adopt the term \emph{convention} instead of \emph{a priori assumption}. Indeed, after this confrontation with Schlick we see in Reichenbach's works no longer any explicit use of the Neo-Kantian \emph{a priori} whereas there is ample reference to conventions. So it appears that Reichenbach has made a significant shift: going from \emph{a priori necessity} of concepts to concepts that are \emph{arbitrary conventions}.

However, at the same time this standard story already suggests a line of continuity: Reichenbach, under Schlick's pressure, changed labels. The relativised a priori becomes a `convention', as a result of Schlick's argument that the Reichenbachian relativised a priori is not really a Kantian a priori at all and that, moreover, it is bad politics to associate oneself with the old-fashioned and rigid philosophy of Kant. Clearly, this argument of Schlick's is directed against the name and its connotations, and not so much against the content and use of Reichenbach's a priori.

Indeed, the terminological concession made by Reichenbach hides a significant difference between fully arbitrary stipulations and the conventions Reichenbach appealed to after 1920.\footnote{See (\cite{dieks2010}) for a more extensive discussion.} In particular, Reichenbach remains faithful to his ideas about continuous extensions of earlier theories---even when he occasionally pays lip service to the idea that we are dealing with "definitions that like all definitions are completely arbitrary". This observation explains, for example, Reichenbach's wrestling  with the `conventionality' of geometry (\cite{r1928}): on the one hand he stresses the logical possibility to employ any geometry one likes, on the other he proposes his famous methodological rule to set universal forces to zero, in order to restrict arbitrariness, fix the geometry uniquely, and stay close to previous theories. It seems, in this case and other examples, that a tension enters Reichenbach's work originating from his continued thinking in terms of ideas very close to his relativised a priori, while speaking about this in terms of conventions (\cite{dieks2010}).

\section{Conclusion}

The concept of probability, on which Reichenbach wrote his 1916 dissertation, turns out to play a pivotal role in Reichenbach's philosophy; in fact, we can understand Reichenbach's later logical positivist views to a large extent by starting from his earliest Neo-Kantian musings about probability.
The story starts with the central problems of the classical interpretation of probability: the connections between equality of chances, equipossibility, and physical symmetries. That physical symmetries lead to equipossibility and then on to equality of chances might seem a self-evident line of reasoning, but for someone unacquainted with this inference from physical properties to equipossibility, the appearance of chances remains a mystery.
Reichenbach's principle of the existence of a continuous probability function was born in the context of these classical problems. The principle makes it possible to justify statements about the equality of chances, so that the notion that physical symmetries lead to equipossibility no longer need to be taken as primitive. Reichenbach shows, following Poincar\'{e}, that if we assume the existence of a continuous probability function, then the assignment of equal probabilities to events ``that differ little from each other'' follows naturally---and this in turn makes it possible to compute other probabilities as well.


In his dissertation Reichenbach motivated the introduction of the principle of the existence of a probability function in Kantian fashion, producing a transcendental deduction of it. He argued that this principle constituted a natural extension of the Kantian a priori scheme, needed in order to make sense of experience even in the presence of arbitrary fluctuations in our measurement results. Going further, in 1920 Reichenbach emphasized the indispensable role this principle plays in a broader Neo-Kantian context: he showed how the introduction of probability considerations makes it possible to adjust Kantian epistemology to the theories of relativity by the procedure of \emph{gradual extension}.

In 1949 Reichenbach had left behind Kant's terminology entirely. His publications from the early twenties onwards certainly suggest a turn away from Kantianism and towards logical positivism. However, as we have argued, underneath this apparent drastic change in Reichenbach's philosophy there is an underlying continuity that makes his epistemology into a coherent whole. In particular, Reichenbach's famous `pragmatic solution of the problem of induction' appears as a direct continuation of his early thoughts about the justification of probability statements.


The argument presented here, therefore, is that the influence of Kantian thought on Reichenbach's epistemology has not vanished in his later work, in particular his 1949 and 1935 books, and has left clearly recognizable traces in this work. Could this perspective of continuity also be turned around? Could one maintain that it is not Reichenbach's later work that shows Kantian traces, but rather his earlier work that already betrays logical positivist leanings? That is not plausible: as we have seen, Reichenbach's later ideas deviate subtly though significantly from those of his logical positivist and frequentist contemporaries, most strikingly by the (conditional) a priori commitments they contain. Adding this to the undisputed and self-avowed Kantian character of Reichenbach's earlier work, it is safe to conclude that these Kantian influences have kept their force during the whole of Reichenbach's career.



\end{document}